\documentclass[article,onecolumn]{IEEEtran}
\usepackage{amsfonts, epsfig, amssymb, amsmath}
\usepackage[normalem]{ulem}
\usepackage{color}
\newcommand{\ol}{\setlength{\itemsep}{0pt.}\begin{enumerate}}
\newcommand{\eol}{\end{enumerate}\setlength{\itemsep}{-\parsep}}
\newcommand{\ignore}[1]{}
\title{An upper bound on $\ell_q$ norms of noisy functions}
\author{Alex Samorodnitsky\thanks{School of Engineering and Computer Science,
The Hebrew University of Jerusalem,
Jerusalem 91904, Israel. Research partially supported by ISF
grant 1724/15.}}

\begin{document}
\date{}
\maketitle


\newtheorem{THEOREM}{Theorem}[section]
\newenvironment{theorem}{\begin{THEOREM} \hspace{-.85em} {\bf :}
}%
                        {\end{THEOREM}}
\newtheorem{LEMMA}[THEOREM]{Lemma}
\newenvironment{lemma}{\begin{LEMMA} \hspace{-.85em} {\bf :} }%
                      {\end{LEMMA}}
\newtheorem{COROLLARY}[THEOREM]{Corollary}
\newenvironment{corollary}{\begin{COROLLARY} \hspace{-.85em} {\bf
:} }%
                          {\end{COROLLARY}}
\newtheorem{PROPOSITION}[THEOREM]{Proposition}
\newenvironment{proposition}{\begin{PROPOSITION} \hspace{-.85em}
{\bf :} }%
                            {\end{PROPOSITION}}
\newtheorem{DEFINITION}[THEOREM]{Definition}
\newenvironment{definition}{\begin{DEFINITION} \hspace{-.85em} {\bf
:} \rm}%
                            {\end{DEFINITION}}
\newtheorem{EXAMPLE}[THEOREM]{Example}
\newenvironment{example}{\begin{EXAMPLE} \hspace{-.85em} {\bf :}
\rm}%
                            {\end{EXAMPLE}}
\newtheorem{CONJECTURE}[THEOREM]{Conjecture}
\newenvironment{conjecture}{\begin{CONJECTURE} \hspace{-.85em}
{\bf :} \rm}%
                            {\end{CONJECTURE}}
\newtheorem{MAINCONJECTURE}[THEOREM]{Main Conjecture}
\newenvironment{mainconjecture}{\begin{MAINCONJECTURE} \hspace{-.85em}
{\bf :} \rm}%
                            {\end{MAINCONJECTURE}}
\newtheorem{PROBLEM}[THEOREM]{Problem}
\newenvironment{problem}{\begin{PROBLEM} \hspace{-.85em} {\bf :}
\rm}%
                            {\end{PROBLEM}}
\newtheorem{QUESTION}[THEOREM]{Question}
\newenvironment{question}{\begin{QUESTION} \hspace{-.85em} {\bf :}
\rm}%
                            {\end{QUESTION}}
\newtheorem{REMARK}[THEOREM]{Remark}
\newenvironment{remark}{\begin{REMARK} \hspace{-.85em} {\bf :}
\rm}%
                            {\end{REMARK}}

\newcommand{\thm}{\begin{theorem}}
\newcommand{\lem}{\begin{lemma}}
\newcommand{\pro}{\begin{proposition}}
\newcommand{\dfn}{\begin{definition}}
\newcommand{\rem}{\begin{remark}}
\newcommand{\xam}{\begin{example}}
\newcommand{\cnj}{\begin{conjecture}}
\newcommand{\mcnj}{\begin{mainconjecture}}
\newcommand{\prb}{\begin{problem}}
\newcommand{\que}{\begin{question}}
\newcommand{\cor}{\begin{corollary}}
\newcommand{\prf}{\noindent{\bf Proof:} }
\newcommand{\ethm}{\end{theorem}}
\newcommand{\elem}{\end{lemma}}
\newcommand{\epro}{\end{proposition}}
\newcommand{\edfn}{\bbox\end{definition}}
\newcommand{\erem}{\bbox\end{remark}}
\newcommand{\exam}{\bbox\end{example}}
\newcommand{\ecnj}{\bbox\end{conjecture}}
\newcommand{\emcnj}{\bbox\end{mainconjecture}}
\newcommand{\eprb}{\bbox\end{problem}}
\newcommand{\eque}{\bbox\end{question}}
\newcommand{\ecor}{\end{corollary}}
\newcommand{\eprf}{\bbox}
\newcommand{\beqn}{\begin{equation}}
\newcommand{\eeqn}{\end{equation}}
\newcommand{\wbox}{\mbox{$\sqcap$\llap{$\sqcup$}}}
\newcommand{\bbox}{\vrule height7pt width4pt depth1pt}
\newcommand{\qed}{\bbox}
\def\sup{^}

\def\H{\{0,1\}^n}

\def\S{S(n,w)}

\def\g{g_{\ast}}
\def\xop{x_{\ast}}
\def\y{y_{\ast}}
\def\z{z_{\ast}}

\def\f{\tilde f}

\def\n{\lfloor \frac n2 \rfloor}

\def \E{\mathop{{}\mathbb E}}
\def \R{\mathbb R}
\def \Z{\mathbb Z}
\def \F{\mathbb F}
\def \T{\mathbb T}

\def \x{\textcolor{red}{x}}
\def \r{\textcolor{red}{r}}
\def \Rc{\textcolor{red}{R}}

\def \noi{{\noindent}}

\def \iff{~~~~\Leftrightarrow~~~~}

\def \queq {\quad = \quad}

\def\<{\left<}
\def\>{\right>}
\def \({\left(}
\def \){\right)}

\def \e{\epsilon}
\def \l{\lambda}

\def\Tp{Tchebyshef polynomial}
\def\Tps{TchebysDeto be the maximafine $A(n,d)$ l size of a code with distance $d$hef polynomials}
\newcommand{\rarrow}{\rightarrow}

\newcommand{\larrow}{\leftarrow}

\overfullrule=0pt
\def\setof#1{\lbrace #1 \rbrace}

\begin{abstract}

\noi Let $T_{\e}$, $0 \le \e \le 1/2$, be the noise operator acting on functions on the boolean cube $\H$. Let $f$ be a nonnegative function on $\H$ and let $q \ge 1$. We upper bound the $\ell_q$ norm of $T_{\e} f$ by the average $\ell_q$ norm of conditional expectations of $f$, given sets of roughly $(1-2\e)^{r(q)} \cdot n$ variables, where $r$ is an explicitly defined function of $q$.

\noi We describe some applications for error-correcting codes and for matroids. In particular, we derive an upper bound on the weight distribution of BEC-capacity achieving binary linear codes and their duals. This improves the known bounds on the linear-weight components of the weight distribution of constant rate binary Reed-Muller codes for all (constant) rates.

\end{abstract}

\section{Introduction}

\noi This paper considers the well-known problem of quantifying the decrease in the $\ell_q$ norm of a function on the boolean cube when this function is acted on by the noise operator.

\noi Given a noise parameter $0 \le \e \le 1/2$, the noise operator $T_{\e}$ acts on functions on the boolean cube as follows: for $f:~\H \rarrow \R$, $T_{\e} f$ at a point $x$ is the expected value of $f$ at $y$, where $y$ is a random binary vector whose $i^{\small{th}}$ coordinate is $x_i$ with probability $1-\e$ and $1 - x_i$ with probability $\e$, independently for different coordinates. Namely,
$\(T_{\e} f\)(x) =  \sum_{y \in \H} \e^{|y - x|}  (1-\e)^{n - |y-x|}  f(y)$,
where $|\cdot|$ denotes the Hamming distance. We will write $f_{\e}$ for $T_{\e} f$, for brevity.

\noi Note that $f_{\e}$ is a convex combination of shifted copies of $f$. Hence, the noise operator decreases norms. An effective way to quantify this decrease for $\ell_q$ norms is given by the hypercontractive inequality \cite{Bonami,Gross,Beckner} (see also \cite{O'Donnell}):
\beqn
\label{hypercontractive}
\|f_{\e}\|_q ~\le~ \|f\|_{1 + (q-1)(1-2\e)^2}.
\eeqn

\noi {\it Entropy} provides another example of a convex homogeneous functional on (nonnegative) functions on the boolean cube. For a nonnegative function $f$ let the entropy of $f$ be given by $Ent(f) = \E f \log_2 f -\E f \log_2 \E f$ (where the expectation is taken w.r.t. the uniform measure on $\H$). The entropy of $f$ is closely related to Shannon's entropy of the corresponding distribution $f / \Sigma f$ on $\H$ (see e.g., the discussion in the introduction of \cite{Sam}), and similarly the entropy of $f_{\e}$ is related to Shannon's entropy of the output of a binary symmetric channel with error probability $\e$ on input distributed according to $f / \Sigma f$. The decrease in entropy (or, correspondingly the increase in Shannon's entropy) after noise is quantified in the "Mrs. Gerber's Lemma" \cite{Wyner-Ziv}: $Ent\(f_{\e}\) \le n \E f \cdot \phi\(\frac{Ent(f)}{n \E f}, ~\e\)$, where $\phi$ is an explicitly given function on $[0,1] \times \left[0,1/2\right]$.

\noi There is a well-known connection between the $\ell_q$ norms of a nonnegative function $f$ and its entropy (see e.g., \cite{Cover-Thomas}): Assume, as we may by homogeneity, that $\E f = 1$. Then $Ent(f) = \lim_{q\rarrow 1} \frac{1}{q-1} \log_2 ||f||^q_q$. The quantity $Ent_q(f) = \frac{1}{q-1} \log_2 ||f||^q_q$ is known as the $q^{\small{th}}$ Renyi entropy of $f$ (\cite{Renyi}).

\noi The starting point for this paper is an alternative way to quantify the decrease in entropy after noise given in \cite{Sam}. To state this result, let us introduce some notation. For $0 \le \l \le 1$, let $T \sim \l$ denote a random subset $T$ of $[n]$ in which each element is chosen independently with probability $\l$. Let $\E(f|T)$ be the conditional expectation of $f$ with respect to $T$, that is a function on $\{0,1\}^n$ defined by $\E(f|T)(x) = \E_{y: y_{|T} = x_{|T}} f(y)$. Then (\cite{Sam}):
\beqn
\label{noisy ent}
Ent\(f_{\e}\) ~\le~ \E_{T \sim (1-2\e)^2} Ent\big(\E(f|T)\big).
\eeqn

\noi In light of the connection between entropy and $\ell_q$ norms, it is natural to ask whether (\ref{noisy ent}) can be extended to an inequality between $\ell_q$ norms (equivalently, Renyi entropies). The main result of this paper is a positive answer to this question.

\thm
\label{thm:main}
Let $f$ be a nonnegative function on $\{0,1\}^n$. Then, for any $q > 1$ holds
\beqn
\label{ineq:main}
\log ||f_{\e}||_q \quad \le \quad \E_{T \sim \l} \log ||\E(f|T)||_q,
\eeqn
with $\l = \l(q,\e) = (1-2\e)^r$, where the exponent $r = r(q)$ is given by the following expression:
\beqn
\label{r}
r(q) \queq \left\{\begin{array}{cc} \frac{1}{2\ln 2} \frac{2^{3-q} \(2^{q-1} - 1\)}{q-1} & 1 < q \le 2 \\ \frac{1}{2\ln 2} \frac{q}{q-1} & q \ge 2 \end{array} \right.
\eeqn
\ethm

\noi Note that (\ref{ineq:main}) is in fact an extension of (\ref{noisy ent}), since assuming $\E f = 1$, dividing both sides by $q-1$ and taking the limit as $q \rarrow 1$, recovers (\ref{noisy ent}). The only thing to observe is that $\lim_{q \rarrow 1} r(q) = 2$.

\rem
Since conditional expectation of a function $f$ is a convex combination of shifted copies of $f$, its norm is smaller than that of $f$, and hence (\ref{ineq:main}) measures the decrease in $\ell_q$-norm under noise, similarly to (\ref{hypercontractive}). In fact, the proof of (\ref{ineq:main}) follows the approach of \cite{Gross} to the proof of (\ref{hypercontractive}). In this approach, we view both sides of the corresponding inequality as functions of $\e$ (for a fixed $q$) and compare the derivatives. Since noise operators form a semigroup it suffices to compare the derivatives at zero, and this is done via an appropriate logarithmic Sobolev inequality (see (\ref{ineq:l-Sob})).
\erem

\subsection{Applications}

\noi Let $C \subseteq \H$ be a binary linear code. Let $r_C(\cdot)$ denote the rank function of the binary matroid defined by $C$. That is, $r_C(T)$ is the rank of the column submatrix of a generating matrix of $C$ which contains columns indexed by $T$. Applying (\ref{ineq:main}) to the scaled characteristic function of $C$, that is $f = \frac{2^n}{|C|} \cdot 1_C$, gives the following claim, connecting between the linear structure of $C$ and its behavior under noise.

\pro
\label{pro:linear}
For any $0 \le \l \le 1$ holds
\beqn
\label{ineq:linear}
\lambda n - \E_{T \sim \lambda} r_C(T) ~\ge~ \max_{1 < q < \infty} \frac{1}{q-1} \cdot \log_2 \E f^q_{\e(q)}, \quad \textup{for} \quad \e(q) = \frac{1 - \l^{1/r(q)}}{2}.
\eeqn
\epro

\noi Let us discuss the bound given by (\ref{ineq:linear}). First, it seems useful to extend it to the limiting values of $q$ as well. Let $F(\l,q) = \frac{1}{q-1} \cdot \log_2 \E f^q_{\e(q)}$, for $1 < q < \infty$. Set $F(\l,1) = \lim_{q \rarrow 1} F(\l,q) = Ent\(f_{\frac{1 - \sqrt{\l}}{2}}\)$ and $F(\l,\infty) = \lim_{q \rarrow \infty} F(\l,q) = \log_2 \|f_{\frac{1 - \l^{2\ln 2}}{2}}\|_{\infty}$. Then we get $\lambda n - \E_{T \sim \lambda} r_C(T) \ge \max_{1 \le q \le \infty} F(\l,q)$. For any $\l$, the three values of $F(\l,q)$ which seem to be easier to understand are $F(\l,1)$, $F(\l,\infty)$ and $F(\l,2) = \log_2 \E f^2_{\frac{1 - \l^{\ln 2}}{2}}$. While simple examples show that $F(\l,1)$ can be both smaller and greater than $F(\l,2)$, intriguingly we always have $F(\l,2) = F(\l,\infty)$, and moreover this quantity is easily expressible in terms of the weight distribution of the code $C$ or of the dual code $C^{\perp}$.

\lem
\label{lem:2=infty}
Let $\(a_0,...,a_n\)$ be the weight distribution of $C$. That is $a_k = \Big |\left\{x \in C, |x| = k\right\} \Big |$, for $k = 0,...,n$. Similarly, let $\(b_0,...,b_n\)$ be the weight distribution of $C^{\perp}$. Then, writing $\theta$ for $\l^{2 \ln 2}$, we have 
\[
F(\l,2) ~=~ F(\l,\infty) ~=~ \log_2 \sum_{i=0}^n b_i \theta^i ~=~ \log_2 \(\frac{1}{|C|} \cdot \sum_{k=0}^n a_k (1-\theta)^k (1+\theta)^{n-k}\).
\]
\elem

\noi We proceed to discuss some applications of (\ref{ineq:linear}).

\subsubsection{Codes achieving capacity for the binary erasure channel}

\noi This is a family of codes for which we can effectively upper bound the LHS of (\ref{ineq:linear}), for $\l$ close to the rate of the code. Recall that the rate of a linear code $R(C)$ is defined as  $\frac{1}{n} \log_2(|C|) = \frac{\textup{dim}(C)}{n}$. A code $C \subseteq \H$ (more precisely a family of codes indexed by $n$) achieves capacity in the block-error rate sense for the BEC (binary erasure channel) if there exist two functions $\delta_1, \delta_2$ which go to zero with $n$ such that for $\l = R(C) + \delta_1$ the probability of decoding error given by $p_e = \textup{Pr}_{T \sim \lambda}\left\{r_C(T) < R(C) \cdot n\right\}$ is upperbounded by $\delta_2$. This immediately implies (writing $R$ for $R(C)$) that
\[
\lambda n - \E_{T \sim \lambda} r_C(T) ~\le~ \lambda n - \(1-\delta_2\) \cdot Rn ~=~ \delta_1 n + \delta_2  Rn ~=~ o(n).
\]

\noi Hence we have the following corollary of (\ref{ineq:linear}).

\cor
\label{cor:BEC-capacity-achieving}
Let $C$ be a BEC capacity achieving code of rate $R$. Then there exists $\l = R + o(1)$ such that for all $1 \le q \le \infty$ holds $F(\l,q) \le o(n)$.
\ecor

\noi The inequality $Ent\(f_{\frac{1 - \sqrt{\l}}{2}}\) = F(\l,1) \le o(n)$ has been interpreted (\cite{OPU}, using (\ref{noisy ent})) as indicating that $C$ is 'well-dispersed' in $\H$ in the sense that the output of the binary symmetric channel with error probability $\e = \frac{1 - \sqrt{\l}}{2} \approx \frac{1 - \sqrt{R}}{2}$ whose input is a random codeword from $C$ is very close in the Kullback-Leibler distance to the uniform distribution. We observe that Corollary~\ref{cor:BEC-capacity-achieving} can be interpreted as providing increasingly stronger measures of proximity of the channel output to uniform, as noise increases beyond $\frac{1 - \sqrt{R}}{2}$.

\noi The inequality $F(\l,2) \le o(n)$ together with Lemma~\ref{lem:2=infty} provide the following bound on the components of the weight distribution of a code which achieves BEC capacity, or a code whose dual achieves BEC capacity. 

\pro
\label{pro:BEC-capacity}
Let $C$ a linear code of rate $R$. Let $\(a_0,...,a_n\)$ be the weight distribution of $C$. For $0 \le k \le n$, let $k^{\ast} = \min\{k,n-k\}$. Let $\theta = R^{2 \ln 2}$. 

\begin{itemize}

\item If $C^{\perp}$ achieves BEC capacity, then for all $0 \le k \le n$ holds
\[
a_k ~\le~ 2^{o(n)} \cdot \(\frac{1}{1-R}\)^{(2 \ln 2) \cdot k^{\ast}}.
\]

\item If $C$ achieves BEC capacity, then for all $0 \le k \le n$ holds
\[
a_k ~\le~ 2^{o(n)} \cdot \left\{\begin{array}{ccc} \frac{|C|}{(1-\theta)^{k^{\ast}} (1+\theta)^{n-k^{\ast}}} & 0 \le k^{\ast} \le \frac{1 - \theta}{2} \cdot n \\ \frac{{n \choose {k^\ast}} \cdot |C|}{2^n} & \mathrm{otherwise} \end{array} \right.
\]
\end{itemize}

\epro

\noi In particular, since the dual of a Reed-Muller code is a Reed-Muller code, and since Reed-Muller codes achieve BEC capacity \cite{KKMPSU}, both bounds in Proposition~\ref{pro:BEC-capacity} hold for
Reed-Muller codes. The bounds in the literature \cite{ASW}, \cite{KKMPU}, \cite{Sberlo-thesis} seem to focus mostly on Reed-Muller codes of rates close to $0$ or $1$, or on weights which increase sublinearly in $n$, but do extend to all weights and to all rates. Comparing with these bounds, it seems that Proposition~\ref{pro:BEC-capacity} improves the bounds on $\{a_k\}$, for $k$ growing linearly in $n$, for constant-rate Reed-Muller codes of all rates $0 < R < 1$. (See Section~\ref{subsec:bnd-compare} for a quick comparison between the bounds.)

\noi Another implication of Proposition~\ref{pro:BEC-capacity} which seems to be worth drawing attention to is that a linear code $C$ of rate $R$ that achieves BEC capacity has $a_k \le 2^{o(n)} \cdot \frac{|C| {n \choose k}}{2^n}$, for $k \in \frac{1 \pm R^{2 \ln 2}}{2} \cdot n$. In other words, in this range of weights, the weight distribution of $C$ is upper-bounded, up to a relatively small error, by the weight distribution of a typical random code of the same rate.

\subsubsection{Bounds on weight distribution of linear codes}
The following (more general) claim is an immediate corollary of the inequality $\lambda n - \E_{T \sim \lambda} r_C(T) \ge F(\l,2)$ and of Lemma~\ref{lem:2=infty}.
\cor
\label{cor:dist-bounds}
Let $C$ be a binary linear code and let $\(b_0,...,b_n\)$ be the distance distribution of the dual code $C^{\perp}$. Then for any $0 \le \l \le 1$ and for any $0 \le i \le n$ holds
\beqn
\label{ineq:dist-comp}
b_i ~\le~ \lambda^{-(2\ln 2) \cdot i} \cdot 2^{\lambda n - \E_{T \sim \lambda} r_C(T)}.
\eeqn
\ecor

\noi This claim gives upper bounds on the components of the weight distribution of $C^{\perp}$, provided we can upper bound $\lambda n - \E_{T \sim \lambda} r_C(T)$.

\subsubsection{Ranks of random subsets in a binary matroid}

\noi We give a different way to write the inequality $\lambda n - \E_{T \sim \lambda} r_C(T) \ge F(\l,2)$, stating it as a lemma since it requires a (simple) argument.

\lem
\label{lem:q=2-linear-restate}
Let $r_C(\cdot)$ be the rank function of the binary matroid on $\{1,...,n\}$ defined by a generating matrix of a linear code $C$ of length $n$. Let $0 \le p \le 1$ and let $t = p^{\frac{1}{2 \ln 2}}$. Then
\[
\log_2 \E_{S \sim p} \(2^{|S| - r_C(S)}\) \quad \le \quad \E_{T \sim t} \big(|T| - r_C(T)\big).
\]
\elem

\noi Let the dimension of $C$ be $k$. Recall that the Tutte polynomial \cite{BO} of the matroid defined by $C$ is  $T_C(x,y) = \sum_{S \subseteq \{1,...,n\}} (x-1)^{k-r_C(S)} (y-1)^{|S| - r_C(S)}$. It is easy to see that an alternative way to write the inequality of Lemma~\ref{lem:q=2-linear-restate} is
\[
\log_2 \left[ p^k(1-p)^{n-k} T_C\(\frac{1}{p},\frac{1+p}{1-p}\) \right] ~\le~ t^{k+1} (1-t)^{n-k-1} \(\frac{d}{dy} T_C\)\(\frac{1}{t},\frac{1}{1-t}\).
\]

\noi As an immediate implication of Lemma~\ref{lem:q=2-linear-restate} we get the following tail bound.

\cor
\label{cor:tail-matroid}
For any $\Delta \ge 0$ holds
\[
Pr_{S \sim p} \left\{|S| - r_C(S) \ge \E_{T \sim t} \Big(|T| - r_C(T)\Big) + \Delta\right\} ~\le~ 2^{-\Delta}.
\]
\ecor

\rem
\label{rem:tails}
Consider a different approach to obtain tail bounds for the function $|S| - r_C(S)$. Let $f(S) = |S| - r_C(S)$. Then $|f(x) - f(y)| \le \|x-y\|$, for any $x, y \in \H$ (here $\|x-y\|$ stands for the Hamming distance between $x$ and $y$) and hence, by the 'bounded differences' inequality \cite{BLM}, for any $t \ge 0$ holds $Pr_{S \sim p} \left\{f(S) \ge \E_{T \sim p} f(T) + t\right\} \le e^{-2t^2/n}$.

\noi Let $\mu: p \mapsto \E_{S \sim p} f(S)$. By the Margulis-Russo formula \cite{O'Donnell},
$\mu' = \frac 1p \E_{S \sim p} \sum_{i \in S} \(f(S) - f\(S \setminus \{i\}\)\)$. Hence, since $f$ is monotone increasing, $\mu' \ge 0$ and $\mu$ is increasing. Moreover, since $f$ is supermodular (by submodularity of rank), it is easy to see that $\mu$ is convex. In particular, since $\mu(0) = 0$, we have $\mu(t) \ge \frac{t}{p} \mu(p)$ for $t \ge p$. Taking everything into account, we have $Pr_{S \sim p} \left\{f(S) \ge \E_{T \sim t} f(T) + \Delta\right\} \le e^{-\frac{2(\mu(t) - \mu(p) + \Delta)^2}{n}} \le e^{-\frac{2((t-p)\mu(p) + p\Delta)^2}{p^2 n}}$.

\noi Note that this does not seem to recover the bound of Corollary~\ref{cor:tail-matroid} when $\mu(t)$ is small, as in the case of BEC capacity achieving codes.
\erem

\noi To conclude, we consider the special case of graphic matroids. For a graph $G = (V,E)$ with $n$ edges, let $M$ be the matroid on $\{1,...,n\}$ whose independent sets are forests in $G$. This is a binary matroid, and Lemma~\ref{lem:q=2-linear-restate} specializes as follows.

\cor
\label{cor:graphs}
Let $G = (V,E)$ be a graph. For a subset of edges $S \subseteq E$, let $c(S)$ denote the number of the connected components in the subgraph $(V,S)$. Then
\[
\log_2 \E_{S \sim p} \(2^{|S| + c(S)}\)  ~\le~ t |E| + \E_{T \sim t} c(T).
\]
\ecor

\noi This paper is organized as follows. We prove Theorem~\ref{thm:main} in Section~\ref{sec:main}. All the remaining proofs are in Section~\ref{sec:lemmas}.

\section{Proof of Theorem~\ref{thm:main}}
\label{sec:main}

\noi We start with a log-Sobolev-type inequality. Recall that the Dirichlet form ${\cal E}(f,g)$ for functions $f$ and $g$ on the boolean cube is defined by ${\cal E}(f,g) = \E_x \sum_{y \sim x} \Big(f(x) - f(y)\Big) \Big(g(x) - g(y)\Big)$. Here $y \sim x$ means that $x$ and $y$ differ in precisely one coordinate.

\thm
\label{thm:l-Sob}
Let $f$ be a nonnegative function on $\H$. Then for any $q \ge 1$ holds
\beqn
\label{ineq:l-Sob}
{\cal E}\(f^{q-1},f\) ~\ge~ 4 r(q) \E f^q \cdot \(n  \cdot \ln ||f||_q ~-~ \sum_{|T| = n-1} \ln ||\E(f|T)||_q\),
\eeqn
where $r(q)$ is given in (\ref{r}).

\noi Moreover for $1 < q < 2$ equality is attained if and only if $f$ is a constant function, and for $q \ge 2$ if and only if $f$ satisfies the following condition: If $x$ and $y$ differ in precisely one coordinate and if $f(x) \not = f(y)$, then $f$ vanishes at either $x$ or $y$.\footnote{Conditions of equality in (\ref{ineq:l-Sob}) are stated for completeness. They are not used in the proof of Theorem~\ref{thm:main}.}
\ethm

\cor
\label{cor:smaller r}
Let $\delta > 0$ be arbitrarily small. For any nonnegative non-constant function $f$ on $\H$ and for any $q > 1$ holds
\[
{\cal E}\(f^{q-1},f\) > 4 \(r(q) - \delta\) \E f^q \cdot \(n  \cdot \ln ||f||_q ~-~ \sum_{|T| = n-1} \ln ||\E(f|T)||_q\).
\]
\ecor

\noi Theorem~\ref{thm:l-Sob} is proved below. For now we assume that it holds and proceed with the proof of Theorem~\ref{thm:main}. We assume, as we may, that the logarithms on both sides of (\ref{ineq:main}) are the natural logarithms. Observe that the claim of the theorem is immediate for constant functions, since both sides of (\ref{ineq:main}) are $\ln(\E f)$. So we may and will assume that $f$ is not constant (note that this implies $f_{\e}$ is non-constant for all $0 \le \e < 1/2$).

\noi Fix $q > 1$. Let $\delta > 0$. We will prove (\ref{ineq:main}) for $\l(q,\e) = (1-2\e)^{r(q) - \delta}$. Taking $\delta$ to zero will then imply (\ref{ineq:main}) for $\l = (1-2\e)^{r(q)}$ as well.

\noi The proof proceeds by induction on $n$. The claim clearly holds for $n = 0$. Let $n > 0$. We assume that the claim holds for all dimensions smaller than $n$, and show that it holds for $n$ as well. For fixed $q$ and $\delta$ both sides of (\ref{ineq:main}) are functions of $f$ and $\e$. Let ${\cal F}(f,\e)$ denote the LHS and ${\cal G}(f,\e)$ the RHS. Clearly ${\cal F}(f,0) = {\cal G}(f,0) = \ln ||f||_q$. We will argue that if ${\cal F}(f,\e) = {\cal G}(f,\e)$ for some $0 \le \e < 1/2$ then necessarily ${\cal F}'(f,\e) < {\cal G}'(f,\e)$ (here and below the derivatives are taken w.r.t. $\e$). It is easy to see that this implies ${\cal F}(f,\e) \le {\cal G}(f,\e)$, for all $0 \le \e \le 1/2$, which is precisely what we need to show.

\noi Due to the fact that the noise operators form a semigroup under composition: $T_{\rho} \circ T_{\e} = T_{\e + \rho - 2\e \rho}$, it will suffice to compare the derivatives at zero (see (\ref{der:zero-general}) and the argument preceding it). This comparison is done in the following lemma.

\lem
\label{lem:der-zero}
Let $f$ be a nonnegative non-constant function on $\H$. Then
\[
{\cal F}'(f,0) ~<~ {\cal G}'(f,0).
\]
\elem

\prf
As shown in \cite{Gross}, we have ${\cal F}'(f,0) = -\frac{{\cal E}\(f^{q-1},f\)}{2\E f^q}$. We claim that
\[
{\cal G}'(f,0) = -2(r(q) - \delta) \cdot \(n \ln ||f||_q - \sum_{|T| = n-1} \ln ||\E(f|T)||_q\),
\]
which will conclude the proof of the lemma by Corollary~\ref{cor:smaller r}. In fact, note that $\l(q,0) = 1$ and that $\l'(q,0) = -2(r(q) - \delta)$. Hence we have:
\[
{\cal G}'(f,0) = \frac{d}{d\e}_{|\e = 0} \E_{T \sim \l} \ln ||\E(f|T)||_q =
\frac{d}{d\e}_{|\e = 0} \(~\l^n \cdot \ln ||f||_q + \l^{n-1} (1-\l) \cdot \sum_{|T| = n-1} \ln ||\E(f|T)||_q\) =
\]
\[
-2(r(q) - \delta) \cdot \(n  \cdot \ln ||f||_q ~-~ \sum_{|T| = n-1} \ln ||\E(f|T)||_q\),
\]
completing the proof.

\eprf

\noi We can now conclude the proof of Theorem~\ref{thm:main}. Assume that ${\cal F}(f,\e) = {\cal G}(f,\e)$, for some $0 \le \e < 1/2$. Consider the functions ${\cal F}\(f_{\e}, \rho\)$ and ${\cal G}\(f_{\e}, \rho\)$, for $0 \le \rho \le 1/2$. We have
\[
{\cal F}\(f_{\e}, \rho\) ~=~ \ln ||\(f_{\e}\)_{\rho}||_q ~=~ \ln ||f_{\e + (1-2\e) \cdot \rho}||_q ~=~ {\cal F}\big(f, \e + (1-2\e) \cdot \rho\big),
\]
and
\[
{\cal G}\(f_{\e},\rho\) ~=~ \E_{R \sim \l(q,\rho)} \ln ||\E(f_{\e} | R)||_q ~\le~ \E_{R \sim \l(q,\rho)} ~\E_{T \subseteq R, T \sim \l(q,\e)} \ln ||\E(f | T)||_q ~=~
\]
\[
\E_{T \sim \l(q,\rho) \cdot \l(q,\e)} \ln ||\E(f | T)||_q ~=~ \E_{T \sim \l\(q,\e + (1-2\e) \cdot \rho\)} \ln ||\E(f | T)||_q ~=~ {\cal G}\big(f,\e + (1-2\e) \cdot \rho\big).
\]
The inequality follows from the induction hypothesis for $R \subset [n]$ and from the assumption ${\cal F}(f,\e) = {\cal G}(f,\e)$ for $R = [n]$.

\noi Hence, using by Lemma~\ref{lem:der-zero} in the first inequality and the fact that ${\cal G}\(f_{\e},0\) = \ln \|f_{\e}\|_q = {\cal F}\(f,\e\) = {\cal G}\(f,\e\)$ in the second inequality, we have
\beqn
\label{der:zero-general}
{\cal F}'(f,\e) ~=~ \frac{1}{1-2\e} \cdot {\cal F}'\(f_{\e},0\) ~<~ \frac{1}{1-2\e} \cdot {\cal G}'\(f_{\e},0\) ~\le~ {\cal G}'(f,\e),
\eeqn
concluding the proof. 

\eprf

\subsection{Proof of Theorem~\ref{thm:l-Sob}}

\noi We start with the base case $n = 1$. This case is dealt with in the following claim.

\pro
\label{pro:2-point}
Let $g$ be a nonnegative function on a $2$-point space with $\E g = 1$. Then
\beqn
\label{2-point}
{\cal E}\(g^{q-1},g\) ~\ge~ 4 r(q) \E g^q \ln \|g\|_q,
\eeqn
where $r(q)$ is given in (\ref{r}).

\noi Moreover for $1 < q < 2$ equality is attained if and only if $g$ is a constant function, and for $q \ge 2$ if and only if $g$ is a constant function, or a mutiple of a characteristic function of a point.

\epro

\noi Proposition~\ref{pro:2-point} is proved below. Here we assume its validity and  proceed with the proof of Theorem~\ref{thm:l-Sob}. First, we restate the $2$-point inequality (\ref{2-point}) in an equivalent form, with two modifications: replacing $\|g\|_q$ with $\E g^q$ and extending the inequality by homogeneity to general nonnegative functions.

\noi Let $g$ be a nonnegative function on a $2$-point space. Then
\[
{\cal E}\(g^{q-1},g\) ~\ge~ \frac{4r(q)}{q} \E g^q \ln \frac{\E g^q}{\(\E g\)^q}.
\]

\noi We can now extend this inequality to general $n$ by the following standard argument. Let $f$ be a nonnegative function on $\H$. Then, denoting by $e_i$ the $i^{\small{th}}$ unit vector, we have
\[
{\cal E}\(f^{q-1},f\) ~=~ \sum_{i=1}^n \E_x \(f^{q-1}(x) - f^{q-1}(x+e_i)\)\(f(x)-f(x+e_i)\).
\]
For $x$ and $i$, let $f_{x,i}$ denote the restriction of $f$ to the $2$-point space $(x, x+e_i)$. Then the RHS is $\sum_{i=1}^n \frac{1}{2^{n-1}} \sum_{x: x_i = 0} {\cal E}\(f_{x,i}^{q-1},f_{x,i}\)$. By the $2$-point inequality, this is at least
\[
\frac{4r(q)}{q} \cdot \sum_{i=1}^n \frac{1}{2^{n-1}} \sum_{x: x_i = 0} \E f^q_{x.i} \ln \frac{\E f^q_{x.i}}{\(\E f_{x,i}\)^q} ~=~  \frac{4r(q)}{q} \cdot \sum_{i=1}^n \frac{1}{2^{n-1}} \sum_{x: x_i = 0} \E f^q_{x.i} (-\ln) \frac{\(\E f_{x,i}\)^q}{\E f^q_{x.i}}
\]
Let $\theta_{x,i} = \frac{\E f^q_{x,i}}{\sum_{y:y_i = 0} \E f^q_{y,i}}$. Then $\sum_{x:x_i=0} \theta_{x,i} = 1$. Note also that $\sum_{y:y_i = 0} \E f^q_{y,i} = 2^{n-1} \cdot \E f^q$. By convexity of $(-\ln)$ the RHS above is at least
\[
\frac{4r(q)}{q} \cdot \E_x f^q \cdot \sum_{i=1}^n  (-\ln)\(\sum_{x:x_i=0} \theta_{x,i} \frac{\(\E f_{x,i}\)^q}{\E f^q_{x,i}}\) ~=~ \frac{4r(q)}{q} \cdot \E_x f^q \cdot \sum_{i=1}^n (-\ln) \frac{\sum_{x:x_i = 0} \(\E f_{x,i}\)^q}{2^{n-1} \cdot \E_x f^q}
\]
\[
=~ \frac{4r(q)}{q} \cdot \E_x f^q \cdot\sum_{i=1}^n \ln \frac{\E f^q}{\E \Big(\E\big(f|[n] \setminus \{i\}\big)^q\Big)} ~=~ 4 r(q) \E f^q \cdot \(n  \cdot \ln ||f||_q ~-~ \sum_{|T| = n-1} \ln ||\E(f|T)||_q\),
\]
proving (\ref{ineq:l-Sob}).

\noi It remains to consider the cases of equality. Tracking back the conditions for equality in the proof above, it is easy to see that equality holds for $f$ if and only if it holds for all one-dimensional restrictions $f_{x,i}$. For $1 < q < 2$ this means that all these restrictions are constant functions, implying $f$ is constant. For $q \ge 2$ this means these restrictions are either constant, or multiples of a characteristic function of a point. Alternatively, if $x$ and $y$ differ in precisely one coordinate and if $f(x) \not = f(y)$, then $f$ vanishes at either $x$ or $y$.
\eprf

\subsection{Proof of Proposition~\ref{pro:2-point}}

\noi There are two cases to consider, $1 < q < 2$ and $q \ge 2$. These cases are dealt with in the following subsections, starting with the somewhat easier case $q \ge 2$.

\subsubsection{The case $q \ge 2$}

\noi We will show that for a nonnegative function $g$ on a $2$-point space with $\E g = 1$ holds
\beqn
\label{q>2}
{\cal E}\(g^{q-1},g\) ~\ge~ \frac{2}{\ln 2} \frac{1}{q-1} \cdot \E g^q \ln \E g^q.
\eeqn

\noi Observe that equality holds in two cases: if $g$ is a constant-$1$ function, a multiple of a characteristic function of a point. We will show that these are the only two cases for which equality holds.

\noi Assume, w.l.o.g., that $g(0) \le g(1)$. Let $0 \le x \le 1$, and let $g(0) = x$ and $g(1) = 2-x$. Then (\ref{q>2}) becomes:
\[
\((2-x)^{q-1} - x^{q-1}\) \cdot \big((2-x) - x\big) \quad \ge \quad \frac{2}{\ln 2} \cdot \frac{1}{q-1} \cdot \frac{x^q + (2-x)^q}{2} \ln\(\frac{x^q + (2-x)^q}{2}\)
\]

\noi Setting $y = x(2-x)^{q-1} + x^{q-1}(2-x)$, $z = x^q + (2-x)^q$, and rearranging, this is the same as
$1 - \frac yz \ge \frac{1}{q-1} \cdot \log_2 \frac z2$. Note that since $x=0$ is one of the cases for which equality holds, we may assume that $x > 0$. Now, let $t = \frac{2-x}{x}$. Then $t \ge 1$, $y = 2^q \cdot \frac{t^{q-1} + t}{(t+1)^q}$, and $z = 2^q \cdot \frac{t^q + 1}{(t+1)^q}$.
Rearranging and simpifying, (\ref{q>2}) becomes
\[
\log_2 \(\(\frac{(t+1)^q}{t^q + 1}\)^{\frac{1}{q-1}}\) ~\ge~ \frac{t^{q-1} + t}{t^q + 1}, \quad \mbox{for} \quad t \ge 1, ~q \ge 2.
\]

\noi Let $s = \(\frac{(t+1)^q}{t^q + 1}\)^{\frac{1}{q-1}}$. Clearly $s > 1$. By convexity of the function $a \rarrow a^q$, we have $s \le 2$. By concavity of logarithm $\log_2(s) \ge s - 1$. (We remark that the only case in which equality holds in this inequality is if $s = 2$, that is $t = 1$, and this implies $g$ is constant.) Hence, it suffices to prove
$\(\frac{(t+1)^q}{t^q + 1}\)^{\frac{1}{q-1}} \ge  \frac{(t+1)\(t^{q-1} + 1\)}{t^q + 1}$.
Simplifying and rearranging, this is the same as $(t+1) \cdot \(t^q + 1\)^{q-2} \ge \(t^{q-1} + 1\)^{q-1}$.

\noi Let $h$ be a function on the two-point space with values $t$ and $1$. Then the inequality above is the same as $\|h\|_1 \cdot \|h\|_q^{q(q-2)} \ge \|h\|_{q-1}^{(q-1)^2}$. This can be shown as follows, using H\"older's inequality in the third step,
\[
\|h\|_{q-1}^{q-1} ~=~ \E h^{q-1} ~=~ \<h^{1/(q-1)}, h^{q(q-2)/(q-1)}\> ~\le~
\]
\[
\| h^{1/(q-1)}\|_{q-1} \cdot \|h^{q(q-2)/(q-1)}\|_{(q-1)/(q-2)} ~=~
\|h\|_1^{1/(q-1)}  \cdot \|h\|_q^{q(q-2)/(q-1)}.
\]

\noi This completes the proof of (\ref{q>2}). Tracing back the conditions for equality in (\ref{q>2}), it is easy to see that it holds only in the two cases mentioned above. \eprf

\subsubsection{The case $1 < q < 2$}

\noi We will show that
\beqn
\label{q<2}
{\cal E}\(g^{q-1},g\) ~\ge~ \frac{2}{\ln 2} \cdot r(q) \cdot \E g^q \ln \E g^q, \quad r(q) = \frac{2^{3-q}\(2^{q-1}-1\)}{q(q-1)},
\eeqn
and that equality holds iff $g$ is a constant-$1$ function.

\noi Let us first deal with the case $g(0) = 0$. In this case, the inequality reduces to $q \ge 2^{3-q} \(2^{q-1}-1\) = 4 - 2^{3-q}$, and it is easy to verify that this is a strict inequality for $1 < q < 2$.

\noi Proceeding as in the proof of the first case, using the same notation, and writing $r$ for $r(q)$, the inequality (\ref{q<2}) is equivalent to $1 - \frac yz \ge r \cdot \log_2 \frac z2$. Substituting $t = \frac{2-x}{x}$ (note that we may assume $x = g(0) > 0$) and rearranging, this is the same as
\[
r \cdot \log_2\(\frac{(t+1)^q}{t^q + 1}\) ~\ge~ \frac{t+t^{q-1}}{t^q + 1} + r \cdot (q-1) - 1.
\]

\noi Since $1 < \frac{(t+1)^q}{t^q + 1} \le 2^{q-1}$, concavity of the logarithm implies $\log_2\(\frac{(t+1)^q}{t^q + 1}\) \ge \frac{q-1}{2^{q-1} - 1} \cdot \(\frac{(t+1)^q}{t^q + 1} - 1\)$. (Equality holds only if $\frac{(t+1)^q}{t^q + 1} = 2^{q-1}$, that is $t =1$, which means that $g$ is constant.) So, it would suffice to show that
$
r \cdot  \frac{q-1}{2^{q-1} - 1} \(\frac{(t+1)^q}{t^q + 1} - 1\) \ge  \frac{t+t^{q-1}}{t^q + 1} + r \cdot (q-1) - 1
$,
which is the same as
$
\frac{r(q-1)}{2^{q-1} - 1} \frac{(t+1)^q}{t^q + 1} \ge \frac{t+t^{q-1}}{t^q + 1} + \frac{r(q-1)2^{q-1} - 2^{q-1}+1}{2^{q-1} - 1}
$.

\noi Multiplying both sides by $\(2^{q-1} - 1\) \(t^q + 1\)$ gives
\[
r(q-1) (t+1)^q ~\ge~ \(2^{q-1} - 1\) \(t+t^{q-1}\) +  \(r(q-1)2^{q-1} - 2^{q-1}+1\) \(t^q + 1\).
\]

\noi Writing the above as $F(t) \ge G(t) + H(t)$, it is easy to see that $F(1) = G(1) + H(1)$ and that $F'(1) = G'(1) + H'(1)$. So it suffices to show $F''(t) \ge G''(t) + H''(t)$, which amounts to (after some simplification and rearranging):
\[
rq(q-1)(t+1)^{q-2} + 2^{q-1}t^{q-3} \cdot \((2-q) + \Big(q - rq(q-1)\Big) \cdot t\) ~\ge~ (2-q)t^{q-3} + qt^{q-2},
\]
or
\[
rq(q-1) \(\frac{t+1}{t}\)^{q-3} (t+1) + 2^{q-1}\cdot \((2-q) + \Big(q - rq(q-1)\Big) \cdot t\) ~\ge~ (2-q) + qt.
\]

\noi Substituting $y = \frac{t+1}{t}$ and rearranging, this is the same as proving for $1 < y \le 2$ that
\[
rq(q-1) y^{q-2} + (2-q)\(2^{q-1}-1\) y ~\ge~ \Big(rq(q-1) - (2q-2)\Big)2^{q-1} + (2q-2).
\]

\noi Denote the LHS by $h(y)$. Then $h'$ is proportional to $\(2^{q-1} - 1\) - rq(q-1) y^{q-3}$. Substituting $r = r(q) = \frac{2^{3-q}\(2^{q-1}-1\)}{q(q-1)}$, we see that $h'$ is proportional to $1 - \(\frac{2}{y}\)^{3-q}$. Hence $h' < 0$ for all $y < 2$, and it suffices to verify the last inequality for $y = 2$. Substituting $y=2$ we get an identity in $q$.

\noi This completes the proof of (\ref{q<2}). Tracing back the conditions for equality, it is easy to see that it holds only if $g$ is a constant function. \eprf

\section{Remaining proofs}
\label{sec:lemmas}

\subsubsection*{Proof of Proposition~\ref{pro:linear}}

\noi We will need some facts from Fourier analysis on the boolean cube \cite{O'Donnell}. First, recall that for a function $f$ on $\H$ and for a subset $T \subseteq [n]$ holds $\E\(f|T\) = \sum_{R \subseteq T} \widehat{f}(R) w_R$, where $\{w_R\}$ are the Walsh-Fourier characters. Recall also that if $f = \frac{2^n}{|C|} \cdot 1_C$, where $C$ is a linear code, then $\widehat{f} = 1_{C^\perp}$.

\noi We also recall a fact from linear algebra. For $T \subseteq [n]$, let $C^{\perp}_T$ be the subspace $\left\{R \subseteq T, R \in C^{\perp}\right\}$. It is well known (and easy to see) that $\textup{dim}~ C^{\perp}_T = |T| - r_C(T)$, where $r_C(\cdot)$ is the dimension of the subset of columns indexed by $T$ in a generating matrix of $C$. Hence, for $x \in \H$:
\[
\E\(f|T\)(x) ~=~ \sum_{R \in C^{\perp}_T} w_R(x) ~=~ \left\{\begin{array}{ccc} |C^{\perp}_T| =2^{|T| - r_C(T)} & \mbox{if} & x \in \(C^{\perp}_T\)^{\perp} \\ 0 & \mbox{otherwise} \end{array} \right.
\]

\noi And hence, $\|\E\(f|T\)\|^q_q = \E\(\E\(f|T\)^q\) = 2^{(q-1) (|T| - r_C(T))}$. Using this in (\ref{ineq:main}) gives, for any $q > 1$ that
\[
\log_2 \E f^q_{\e} ~\le~ (q-1) \E_{T \sim \lambda} (|T| - r_C(T)) ~=~ (q-1) \cdot \(\lambda n - \E_{T \sim \lambda} r_C(T)\),
\]
where $\lambda = (1-2\e)^{r(q)}$. Rearranging, we get the claim of the proposition. \eprf

\subsubsection*{Proof or Lemma~\ref{lem:2=infty}}

\noi Recall that for a function $f$ on $\H$ and for $0 \le \e \le 1/2$ holds $\widehat{f_{\e}}(R) = \widehat{f}(R) \cdot (1-2\e)^{|R|}$. Let $\(b_0,...,b_n\)$ be the distance distribution of $C^{\perp}$. Then, since $\e(2) = \frac{1 - \l^{\ln 2}}{2}$, we have
\[
2^{F(\l,2)} ~=~ \sum_{R \in \H} \widehat{f}^2(R) \cdot (1-2\e)^{2|R|} ~=~ \sum_{i=0}^n b_i \l^{(2\ln 2) \cdot i} ~=~ \sum_{i=0}^n b_i \theta^i.
\]

\noi On the other hand, it is easy to see that for $f = \frac{2^n}{|C|} \cdot 1_C$ and for any $0 \le \e \le 1/2$ holds $\|f_{\e}\|_{\infty} = \|\sum_{R \in \H} \widehat{f}(R) (1-2\e)^{|R|} w_R\|_{\infty} = \sum_{R \in \H} \widehat{f}(R) (1-2\e)^{|R|} = f_{\e}(0)$ (since $\widehat{f}$ is nonnegative).
Hence, recalling $\e(\infty) = \frac{1 - \l^{2\ln 2}}{2}$,
\[
2^{F(\l,\infty)} ~=~ \|f_{\e}\|_{\infty} ~=~ f_{\e}(0) ~=~ \sum_{R \in \H} \widehat{f}(R) \cdot (1-2\e)^{|R|} ~=~ \sum_{i=0}^n b_i \l^{(2\ln 2) \cdot i} ~=~ \sum_{i=0}^n b_i \theta^i.
\]

\noi Next, recall that for the function $g(x) = (1-\theta)^{|x|} (1+\theta)^{n - |x|}$ on $\H$ holds $\widehat{g}(R) = \theta^{|R|}$. Hence, by Parseval's identity,
\[
\sum_{i=0}^n b_i \theta^i ~=~ \sum_{R \in \H} \widehat{f}(R) \widehat{g}(R)  ~=~ \frac{1}{2^n} \sum_{x \in \H} f(x) g(x) = \frac{1}{|C|} \cdot \sum_{k=0}^n a_k (1-\theta)^{k} (1+\theta)^{n-k}.
\]
\eprf

\subsubsection*{Proof or Proposition~\ref{pro:BEC-capacity}}

\noi Assume that $C^{\perp}$ achieves BEC capacity. Note that the rate of $C ^{\perp}$ is $1-R$ and that the dual of $C^{\perp}$ is $C$. By Corollary~\ref{cor:BEC-capacity-achieving} and Lemma~\ref{lem:2=infty}, there exists $\l = 1 - R + o(1)$ such that $\sum_{k=0}^n a_k \l^{(2\ln 2)\cdot k} = 2^{o(n)}$. Hence, for all $0 \le k \le n/2$ holds
\[
a_k ~\le~ 2^{o(n)} \cdot \(\frac{1}{\l}\)^{(2\ln 2) \cdot k} ~\le~ 2^{o(n)} \cdot \(\frac{1}{1-R}\)^{(2\ln 2) \cdot k^{\ast}}.
\]

\noi Next, we claim that for any $0 \le \alpha \le 1$ holds $\sum_{k=0}^n a_k \alpha^{n-k} \le \sum_{k=0}^n a_k \alpha^k$. To see this, recall that the Fourier transform on $\H$ is its own inverse, up to normalization, and hence (see the proof of Lemma~\ref{lem:2=infty}) we have for any $\alpha \in \R$ that $\widehat{\alpha^{|x|}}(R) = \frac{1}{2^n} (1-\alpha)^{|R|} (1+\alpha)^{n - |R|}$. Let $f = \frac{2^n}{|C|} \cdot 1_C$. Then, for $0  < \alpha \le 1$, \footnote{Note that for $\alpha = 0$ the claim $\sum_{k=0}^n a_k \alpha^{n-k} \le \sum_{k=0}^n a_k \alpha^k$ is trivial, since $a_0 = 1 \ge a_n$.} we have, by Parseval's identity,
\[
\sum_{k=0}^n a_k \alpha^{n-k} ~=~ \frac{|C|}{2^n} \cdot  \alpha^n \sum_{x \in \H} f(x) \(\frac{1}{\alpha}\)^{|x|} ~=~ \frac{|C|}{2^n} \cdot \sum_{R \in \H} (-1)^{|R|} \widehat{f}(R) (1-\alpha)^{|R|} (1+\alpha)^{n - |R|} ~\le
\]
\[
\frac{|C|}{2^n} \cdot \sum_{R \in \H} \widehat{f}(R) (1-\alpha)^{|R|} (1+\alpha)^{n - |R|} ~=~  \frac{|C|}{2^n} \cdot \sum_{x \in \H} f(x) \alpha^{|x|} ~=~ \sum_{k=0}^n a_k \alpha^k.
\]
For the inequality in the third step, note that $\widehat{f} = 1_{C^{\perp}}$ is nonnegative. 

\noi In particular, setting $\alpha = \l^{2\ln 2}$, we have for all $n/2 < k \le n$ that $a_k ~\le~ 2^{o(n)} \cdot \(\frac{1}{\l}\)^{(2\ln 2) \cdot (n-k)}~\le~2^{o(n)} \cdot \(\frac{1}{1-R}\)^{(2\ln 2) \cdot k^{\ast}}$. This completes the proof of the first claim of the proposition.

\noi  Assume now that $C$ achieves BEC capacity. Let $\(b_0,...,b_n\)$ be the weight distribution of $C^{\perp}$. By Corollary~\ref{cor:BEC-capacity-achieving} and Lemma~\ref{lem:2=infty}, for $\theta = R^{2 \ln 2}$ holds $\sum_{i=0}^n b_i \theta^i = 2^{o(n)}$. Clearly this also holds if we replace $\theta$ with any $0 \le \alpha \le \theta$. Using Parseval's identity as in the proof of Lemma~\ref{lem:2=infty}, we have that $\sum_{k=0}^n a_k (1-\alpha)^k (1 + \alpha)^{n-k} \le 2^{o(n)} \cdot |C|$, for any $0 \le \alpha \le \theta$.

\noi Hence, for all $0 \le k \le n/2$ holds $a_k \le 2^{o(n)} \cdot \frac{|C|}{\max_{0 \le \alpha \le \theta} \left\{(1-\alpha)^k (1 + \alpha)^{n-k}\right\}}$. Let $H(x) = x \log_2\(\frac 1x\) + (1-x) \log_2\(\frac{1}{1-x}\)$ be the binary entropy function. It is easy to see that
\[
\max_{0 \le \alpha \le \theta} \left\{(1-\alpha)^k (1 + \alpha)^{n-k}\right\} ~=~ \left\{\begin{array}{ccc} (1-\theta)^k (1 + \theta)^{n-k} & \mathrm{if} & 0 \le k \le \frac{1 - \theta}{2} \cdot n \\ 2^{\(1-H\(k/n\)\) \cdot n} & \mathrm{if} & \frac{1 - \theta}{2} \cdot n \le k \le n/2 \end{array}\right..
\]

\noi Recalling that, by Stirling's formula, for any $0 \le k \le n$ holds $2^{H\(k/n\) \cdot n} \le O\(\sqrt{n}\) \cdot {n \choose k}$, this proves the second claim of the proposition for $0 \le k \le n/2$. 

\noi To deal with the complementary case $n/2 < k \le n$, observe that $\sum_{k=0}^n a_k (1-\alpha)^k (1 + \alpha)^{n-k} = (1+\alpha)^n \cdot \sum_{k=0}^n a_k \(\frac{1 - \alpha}{1 + \alpha}\)^k \ge (1+\alpha)^n \cdot \sum_{k=0}^n a_k \(\frac{1 - \alpha}{1 + \alpha}\)^{n-k} = \sum_{k=0}^n a_k (1-\alpha)^{n-k} (1 + \alpha)^k$.
\eprf

\subsubsection*{Proof or Lemma~\ref{lem:q=2-linear-restate}}

\noi Let $\e = \frac{1 - \sqrt{p}}{2}$. Using (\ref{ineq:main}) with $q = 2$ we have, for $f = \frac{2^n}{|C|} \cdot 1_C$, that
\[
\E f^2_{\e} ~=~ \sum_{S \in \H} \widehat{f}^2(S) (1-2\e)^{2|S|} ~=~ \sum_{S \in \H} \widehat{f}^2(S) p^{|S|} ~=~ \E_{S \sim p} \E \({\E}^2(f|S)\) ~=~ \E_{S \sim p} 2^{|S| - r_C(S)}.
\]

\noi Hence the claim of the lemma follows from the inequality $\lambda n - \E_{T \sim \lambda} r_C(T) \ge F(\l,2)$ with $\l = (1-2\e)^{r(2)} = (1-2\e)^{\frac{1}{\ln 2}} = p^{\frac{1}{2\ln 2}}$. \eprf

\subsubsection*{Proof or Corollary~\ref{cor:graphs}}

\noi For $S \subseteq E$, the matroid rank of $S$ is given by $r(S) = \sum_{i=1}^{c(S)} \(|C_i| -1 \) = |V| - c(S)$, where $\{C_i\}$ are the connected components of $(V,S)$. Hence, $|S| - r(S) = |S| + c(S) - |V|$. Substituting this into the claim of Lemma~\ref{lem:q=2-linear-restate} gives the claim of this corollary.
\eprf

\subsubsection{Comparing the bounds on the weight distribution of Reed-Muller codes}
\label{subsec:bnd-compare}

\noi As far as we know, the best known bounds in the literature are due to \cite{Sberlo-thesis} (Theorem 1.3). They state, in the notation of Proposition~\ref{pro:BEC-capacity}, that for the Reed-Muller code of rate $R$ and for $k$ linear in $n$ we have
\beqn
\label{bnd-SbSp}
a_k ~\le~ 2^{CRk^{\ast}\(2 \log_2\(\frac{n}{k^{\ast}}\) + 3\)},
\eeqn
where $C$ is an absolute constant, $C \ge 30$. We proceed to compare (\ref{bnd-SbSp}) with the bounds in Proposition~\ref{pro:BEC-capacity}. We ignore the $2^{o(n)}$-factor in these bounds, since it is negligible for $0 < R < 1$ and $k$ linear in $n$.

\noi Replacing the term $\log_2\(\frac{n}{k^{\ast}}\)$ in (\ref{bnd-SbSp}) by $1$ (which is a smaller quantity, since $k^{\ast} \le n/2$), it is easy to see that (\ref{bnd-SbSp}) is larger than the first bound of Proposition~\ref{pro:BEC-capacity} unless $R \ge 1 - \delta$, for a very small $\delta$ (specifically, $\delta < 2^{-100}$). Assume that $R = 1 - \e$, where $\e$ is very small. Consider the second bound of Proposition~\ref{pro:BEC-capacity}. If $k^{\ast} \ge \frac{1 - R^{2 \ln 2}}{2} \cdot n$, then this bound is at most ${n \choose {k^{\ast}}}$, and it is easy to see that it is smaller than (\ref{bnd-SbSp}). For $0 < k^{\ast} \le \frac{1 - R^{2 \ln 2}}{2} \cdot n$, the term $\log_2\(\frac{n}{k^{\ast}}\)$ in (\ref{bnd-SbSp}) behaves like $\log_2\(\frac{1}{\e}\)$, and hence (\ref{bnd-SbSp}) behaves like $\(\frac{1}{\e}\)^{C_1 k^{\ast}}$, for $C_1 \ge 60$, and this is larger than the first bound in Proposition~\ref{pro:BEC-capacity}, which is  $\(\frac{1}{\e}\)^{(2 \ln 2) \cdot k^{\ast}}$. \eprf

\section*{Acknowledgments}

\noi I am grateful to Or Ordentlich for many very helpful conversations and valuable remarks. I would also like to thank the anonymous referees for their numerous and very useful comments.

\begin{IEEEbiographynophoto}{Alex Samorodnitsky}
received the B.A.\ degree in
computer science and mathematics (1987), the M.Sc.\
degree in mathematics (1990), and the Ph.D.\ degree in mathematics (1998), all from the Hebrew University of Jerusalem.

He is now a Professor of Computer Science at the School of Engineering and Computer Science in the Hebrew University, which he joined in 2001 as an Assistant Professor. His research interests are in theoretical computer science, coding theory, and combinatorics.

\end{IEEEbiographynophoto}
\end{document}